# Phonon confinement and substitutional disorder in $Cd_{1-x}Zn_xS$ Nanocrystals

Satyaprakash Sahoo*, S. Dhara, V. Sivasubramanian, S. Kalavathi and A. K. Arora

Materials Science Division, Indira Gandhi Centre for Atomic Research,

Kalpakkam 603102, India

**Abstract**

1-LO optical phonons in free-standing mixed $Cd_{1-x}Zn_xS$ nanocrystals, synthesized using chemical precipitation, are investigated using Raman spectroscopy. As expected for the nanocrystals, the 1-LO modes are found to appear at slightly lower wavenumbers than those in the bulk mixed crystals and exhibit one mode behavior. On the other hand, the line broadening is found to be much more than that can be accounted on the basis of phonon confinement. From the detailed line shape analysis it turns out that the substitutional disorder in the mixed crystals contributes much more to the line broadening than the phonon confinement. The linewidth arising from these mechanisms are also extracted from the analysis.

*Author to whom correspondence should be addressed;

Electronic mail: satyasahoo@igcar.gov.in, satya504@gmail.com

## Introduction

Semiconductor nanoparticles of mixed crystals are of considerable current interest because of the possibility of controlling optical properties by varying the composition as well as by size. These materials have potential applications in many areas that include nonlinear optics and opto-electronic devices.[1-4] $Cd_{1-x}Zn_xS$ semiconductor doped glasses are the most promising materials for long-pass optical filters. Optical and electronic properties of both nanocrystalline[5,6] and bulk[7,8] $Cd_{1-x}Zn_xS$ have been studied by many researchers. Understanding of the fundamental properties of mixed crystal systems is also important from the point of view of applications. For example, the transport properties of semiconductors are essentially controlled by the phonon scattering. The mixed crystals are known to show two type of phonon behavior: (a) one-mode and (b) two-mode behavior.[7] In the one-mode behavior the zone-centre optical phonon frequency varies monotonically with composition from one end member to that of other. For example, this type of behavior is exhibited by $Cd_{1-x}Zn_xS$, $Ca_xSr_{1-x}F_2$ and $Sr_xBa_{1-x}F_2$.[9,10] Whereas in two-mode behavior the zone-centre optical phonons of each of the end members are present at intermediate compositions. $CdS_{1-x}Se_x$, $ZnS_{1-x}Te_x$, $Ga_{1-x}In_xAs$ are found to show a two-mode behavior.[11-13]

Although the phonons in nano-$Cd_{1-x}Zn_xS$ mixed crystals have been reported,[6,14] most of the studies are on semiconductor-doped glass containing $Cd_{1-x}Zn_xS$ nanoparticles. Synthesis of such nanocrystals dispersed in glass matrix involves the melting of the matrix followed by quenching. Because of different thermal expansions of the host and the nanocrystals, the cooling to ambient after synthesis at elevated temperature invariably leads to stressed nanocrystals.[15] Compressive stresses on nanoparticles result in higher phonon wavenumbers.[15,16] On the other hand, the optical phonon confinement in nanoparticles leads to a red-shift of the Raman peak centre.[17] Furthermore, the phonon line shape also exhibits an asymmetric broadening. In addition, substitutional disorder in mixed crystals is also known to result in broad spectra.[18] Thus, the phonon line-shape is influenced by several factors such as (a) stress (b) confinement effect (c) substitutional disorder.

It is not straight forward to separate the contribution of each of these three effects in embedded nanocrystals. There is no report of a systematic study and analysis of Raman line-shape in mixed crystal nanoparticles.

In order to understand and quantify the contributions of various effects to the Raman spectra, it is of interest to examine the phonon line-shapes in the free-standing mixed crystal nanoparticles. Here we report a Raman spectroscopic study of zone-centre longitudinal optical (LO) phonon in cubic-$Cd_{1-x}Zn_xS$ nanocrystalline powders over the complete range of composition. The line width and the peak-centers of the Raman spectra are compared with the reported data on single crystals of $Cd_{1-x}Zn_xS$. The line broadening arising from the phonon confinement and substitutional disorder are estimated from a quantitative fitting of Raman line-shape. The possible reasons for the additional broadening of the spectra in mixed crystals are also discussed.

Experimental

$Cd_{1-x}Zn_xS$ nanocrystals were synthesized at room temperature by chemical precipitation route in aqueous medium. For a typical synthesis of $Cd_{1-x}Zn_xS$ nanoparticle ($x = 0.25$), aqueous solutions of $Cd(CH_3COO)_2$ (0.375 mmol) and $Zn(CH_3COO)_2$ (0.125 mmol) were prepared. While stirring this solution, 25 ml of $Na_2S$ (0.5 mmol) aqueous solution was added slowly which resulted in the precipitation of $Cd_{0.75}Zn_{0.25}S$ particles. The mixed nanocrystals with $x = 0.50$ and $0.75$ were synthesized by suitably changing the molar ratio of the constituent solutions. In order to get nearly same particle size for different compositions all the synthesis conditions were maintained identical. The precipitate was removed after washing with water several times and dried. $Cd_{1-x}Zn_xS$ nanopcrystals were characterized by X-ray powder diffraction using a (STOE) X-ray diffractometer for phase identifications, estimation of lattice parameters and average particle size. For this, powder was placed on Si (911) plate and Cu-K$_\alpha$ radiation was used. The transmission electron microscopy (TEM) of thenanocrystals was carried out using a Philips CM-200 microscope. The Raman scatteringmeasurement were carried out using 532 nm line of a diode-pumped solid-state laser as the excitation and analyzed using a double-monochromator (Jobin-Yvon U1000), equipped with liquid nitrogen cooled CCD detector. Near-resonant Raman scattering were performed using 325 nm line of He-Cd laser as the excitation and analyzed using a double-subtractive triplemonochromator (Jovin-Yvon T64000), equipped with liquid nitrogen cooled CCD detector for recording the spectra.

Results and Discussion

Figure 1 shows the diffraction pattern of as-synthesized $Cd_{1-x}Zn_xS$ mixed crystal for $x$ = 0, 0.25, 0.5, 0.75 and 1.0. The diffraction peaks are broad and their positions show a systematic shift towards higher $2\theta$ with the increase in Zn composition. The X-ray diffraction patterns agree with the JCPDS data for cubic structure for all compositions and the peak broadening suggests the nanocrystalline nature of the samples. The average particle sizes calculated, using the Debye-Scherrer formula, turn out to be 3.2, 3.7, 4.3, 3.7 and 3.4 nm for $x$ =0, 0.25, 0.5, 0.75 and 1.0 respectively. The systematic shift in peak positions suggests that substitution of Zn leads to lattice contraction in the solid solution. The lattice constants have been calculated and Fig. 2 shows the variation of the lattice constant with Zn composition. The linear variation of lattice constant with composition is consistent with Vegard's law. Figure 3 shows the TEM image and the particle size distribution obtained from the analysis of the micrograph for $x$ = 0.5 nanocrystals. The average particle size turned out to be 5.1 nm with a standard deviation of 0.6 nm, in close agreement with that found from XRD analysis.

Figure 4 shows the Raman spectra of $Cd_{1-x}Zn_xS$ mixed crystal nanoparticles for various compositions. The prominent LO phonon modes and relatively less intense second harmonic of 1-LO phonon are found for all compositions. One can also see that the 1-LO phonon peak shifts monotonically to higher wavenumbers as a function of composition $x$. Genzel et al. [19] have theoretically obtained the phonon behavior in bulk hexagonal $Cd_{1-x}Zn_xS$ mixed crystals. The calculation was based on the modified random-element-isodisplacemenT (MREI) model and the difference in crystal structure was not considered. Thus one can use those results for analysis of LO phonon wavenumbers in both cubic and hexagonal structures. Ichimura *et al.* [20] have experimentally obtained the LO phonon wavenumbers in bulk cubic $Cd_{1-x}Zn_xS$ grown on GaAs substrate. The MREI model as well as the results of Ichimura demonstrate that bulk $Cd_{1-x}Zn_xS$ shows a one-mode behavior. The present results on mixed nanocrystals are similar to those reported on the bulk samples suggesting that nanocrystals of cubic $Cd_{1-x}Zn_xS$ also show the one-mode behavior. The variation of the Raman peak position with Zn composition is compared with the results of the MREI model in Fig. 4. One can also see that the Raman peak centre shifts towards lower frequency as compared to its bulk value. The reason for this will be discussed in the next section.

Line-shape analysis and discussions

As mentioned earlier, the phonon line-shape is influenced by several factors such as phonon confinement effect and substitutional disorder. In order to understand the Raman line broadening in nanoparticles of mixed crystal, one has to be careful in analyzing the Raman line-shape and assigning the relative contribution of these effects to total line broadening. In nanocrystals the lattice periodicity is interrupted at the particle surface which leads to relaxation of $q = 0$ selection rule, where $q$ is the scattering vector. Thus the phonon wave function has to decay to a small value close to the boundary. This allows all the $q$s in the Brillouin zone to contribute to Raman intensity and results in shift as well as broadening of Raman peak in nanocrystals.[17] A Gaussian phonon confinement model was used to calculate the Raman line-shape for different compositions. According to this model the intensity of first order Raman scattering for a nanocrystal of diameter $d$ is,[21]

$$I(\omega) = \int \frac{|C(q)|^2}{[\omega - \omega(q)]^2 + (\Gamma_0/2)^2} d^3q ,\qquad(1)$$

where $C(q,d)$ is the Fourier transformation of the confinement function, $\omega(q)$ is the phonon dispersion curve and $\Gamma_0$ is the natural linewidth of zone-center optical phonon in bulk $Cd_{1-x}Zn_xS$. If is the width of the dispersion curve of the LO phonon, then the dispersion curve can be approximated as

$$\omega(q) = \omega_0 - \Delta \sin^2\left(\frac{qa}{2}\right), \qquad (2)$$

where $\Gamma_0$ is the zone-centre LO frequency and $a$ is the lattice parameter. In $Cd_{1-x}Zn_xS$ mixed crystals $\Gamma_0$ varies nearly linearly with composition as shown in Fig. 5. In order to carry out the calculations using Eqns (1) and (2) one also requires $\Delta$ and $\Gamma_0$ for the mixed crystals. However, the phonon dispersion curves have been reported from inelastic neutron scattering measurements only for the end members CdS and ZnS as 35 and 21 cm-1, respectively.[22] In the absence of any neutron scattering data for mixed crystal $Cd_{1-x}Zn_xS$, it may be reasonable to assume that also varies linearly with composition. The natural line widths of the 1-LO phonons in the bulk CdS [23] and that found for bulk ZnS in the present work are nearly the same and have a value around 12 cm$^{-1}$. Hence, in order to quantify the phonon confinement alone, we keep $\Gamma_0$ fixed at 12 cm$^{-1}$ for the mixed crystals also. Thus the confined phonon line-shapes are calculated by taking $\Gamma_0$ as 12 cm$^{-1}$ and $\Delta$ as $\Delta(x) = 35 - 14x$. Table 1 gives the input parameters used for the calculation of Raman line shape for various compositions. It is also important to examine the difference in the line-shape that can arise from the particle size distribution (inhomogeneous broadening) as compared to that calculated from an average size. The effect of particle size distribution on the line-shape can be taken into account as,

$$I(\omega) = \sum_i P(d_i) I(\omega, d_i) \qquad (3)$$

where $P(d_i)$ is the normalized discrete size distribution ($\sum_i P(d_i) = 1$). Figure 6 compares the Raman line shape calculated for $x=0.5$ nanocrystals using the average diameter of 4.3 nm and the actual size distribution (Fig. 3). One can see that the two line shapes differ only marginally. This is because of the narrow size distribution.

Other compositions also show similar result. In view of the nearly same line-shapes, only the average size was used for analysis of Raman spectra.

Figure 7 compares the calculated confined phonon line-shape (dashed curves) with the measured Raman spectra for various compositions in $Cd_{1-x}Zn_xS$. Note that the calculated line-shapes are broad and asymmetric towards low-frequency side. Table 1 also gives the FWHM ($\Gamma_{pc}$) of the calculated spectra representing the broadening arising from the phonon confinement. It is important to point out that the calculated spectra match well with the data only for the end members, i.e., CdS and ZnS, whereas for the intermediate compositions the measured spectra are broader than the calculated one, the disagreement being maximum for $x = 0.5$. The good agreement between the measured and calculated spectra for the end members confirms that in pure CdS and ZnS nanocrystals the phonon confinement is the only mechanism of Raman line broadening. The departure of the calculated Raman line-shape from the data for the intermediate compositions suggests that the substitutional disorder also plays a significant role in determining the line width. We now attempt to quantify the relative contributions of phonon confinement and substitutional disorder to line broadening. In order to obtain reasonable agreement between the calculated and the measured spectra of the mixed nanocrystals the Raman line shape for the intermediate compositions were recalculated using increased natural line widths $\Gamma'_0$ and the calculated profiles are shown as full curves in Fig. 7. The values of $\Gamma'_0$ that yield good agreement are also given in Table 1. Note that these are much higher than the value of 12 $cm^{-1}$ applicable for the pure compounds and represent the broadening arising from substitutional disorder. Figure 8 compares the observed Raman linewidths with those arising from phonon confinement for 3.2 nm (smallest) nanoparticles. The figure also includes $\Gamma'_0$ due to substitutional disorder. One can see that the measured linewidth increases with Zn substitution and reaches a maximum value for $x = 0.5$ and then starts decreasing with further Zn substitution. This is expected because the substitutional disorder is maximum for $x = 0.5$. On the other hand, the line broadening arising from phonon confinement decreases marginally and nearly linearly with the composition. The compositional dependence essentially arises due to $\Delta(x)$. Furthermore, one can see from Table 1 that the contribution from substitutional disorder to the line width is much more than that arising from the phonon confinement. Thus from the present quantitative line-shape analysis it has been possible to extract the contribution of different broadening mechanisms of the LO phonon line-shape in mixed nanocrystals.

A discussion on the microscopic origin of increased intrinsic linewidth in the mixed nanocrystals is in order. The increased linewidth could have contribution from microscopic strains, phonon scattering and bond-length fluctuations.[24,25] Recently a percolation model has been argued to apply for systems exhibiting two-mode behavior. On the other hand, virtual crystal approximation works well for mixed crystals with one-mode behavior. However, both these models attempt to obtain the composition dependence of the phonon wavenumbers; whereas the understanding on the phonon linewidths has remained only qualitative. Larger intrinsic linewidth in the mixed crystal could arise due to the phonon scattering from randomly substituted cations in the lattice as compared to the pure end member crystals. Because of the increased scattering, the phonon lifetime decreases considerably resulting in an increased natural linewidth $\Gamma'_0$ in the mixed crystal. Furthermore, in the $Cd_{1-x}Zn_xS$ mixed crystals the cation-anion bond length ($d_{Cd-S}$ and $d_{Zn-S}$) would fluctuate from site to site depending on the number of next neighbours around a given cation. Although the mixed crystal is predicted to have a single LO-mode frequency under the virtual crystal approximation and MREI model, the existence of a bond-length distribution would result in appearance of a broadened LO mode as found in the present experiment and also reported earlier.[6]

**Conclusion**

Free-standing mixed nanocrystals of $Cd_{1-x}Zn_xS$ were successfully synthesized through aqueous chemical precipitation over the complete range of compositions. Powder XRD reveals the formation of cubic phase for all compositions. Presence of single zone-centre longitudinal optical phonon and its monotonic variation with composition suggests that nanocrystals of $Cd_{1-x}Zn_xS$ show one-mode behavior. The red-shift of the optical phonon wavenumbers in the mixed nanocrystals mainly arise due to phonon confinement. On the other hand, broadening of optical phonon spectra arises due to both phonon confinement and substitutional disorder: the effect of substitutional disorder being more dominant than that of phonon confinement for the intermediate compositions.

**ACKNOWLEDGEMENTS**

Authors thank Dr. R. Mythili for help in obtaining TEM micrographs, Dr. C. S. Sundar for interest in the work, Dr. P. R. Vasudeva Rao for support and Dr. Baldev Raj for encouragement.

Table 1. Different input parameters used for calculating the Raman line-shapes of 1-LOphonon.

| $x$ | $d$ (nm) | $\Delta$ (cm$^{-1}$) | $\Gamma_o$ (cm$^{-1}$) | $\Gamma_{pc}$ (cm$^{-1}$) | $\Gamma_0^{'}$ (cm$^{-1}$) |
|---|---|---|---|---|---|
| 0.0 | 3.2 | 35.0 | 12 | 20.0 | 12 |
| 0.25 | 3.7 | 31.5 | 12 | 17.5 | 26 |
| 0.5 | 4.3 | 28.0 | 12 | 15.5 | 45 |
| 0.75 | 3.7 | 24.5 | 12 | 17.0 | 35 |
| 1.0 | 3.4 | 21.0 | 12 | 18.5 | 12 |

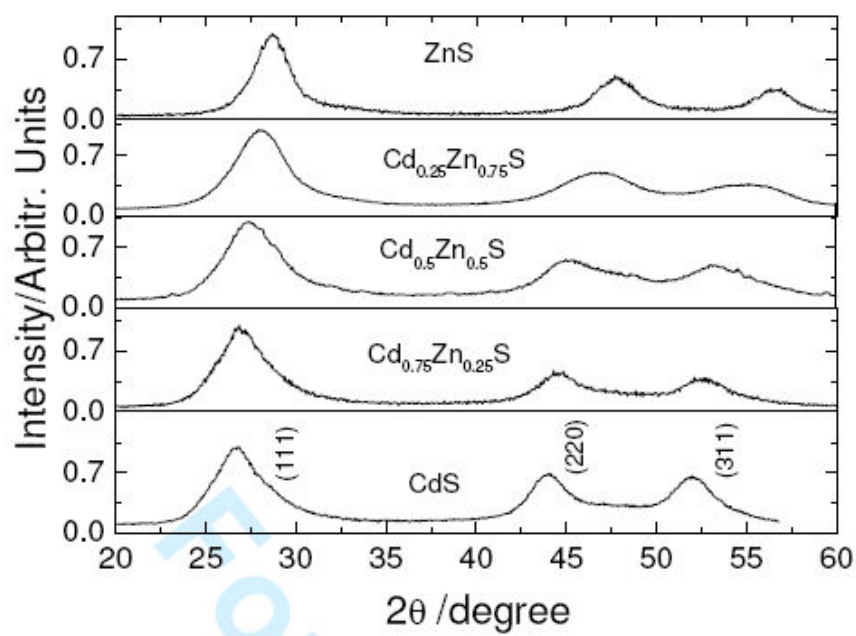

Figure 1. X-ray diffraction patterns of $Cd_{1-x}Zn_xS$ in the region of (111), (220) and (311) reflections.

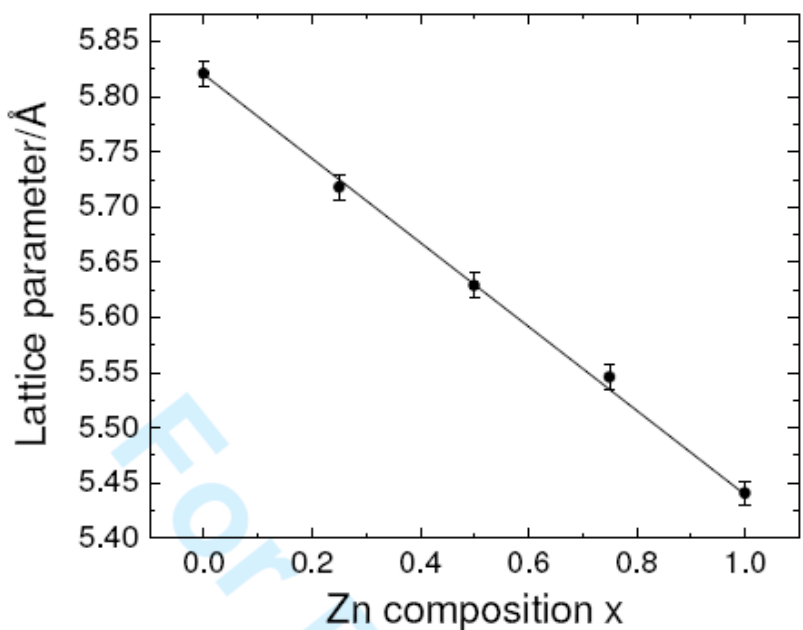

Figure 2. Variation of lattice constant with Zn composition in cubic $Cd_{1-x}Zn_xS$ nanocrystals. Straight line is the liner fit to the data.

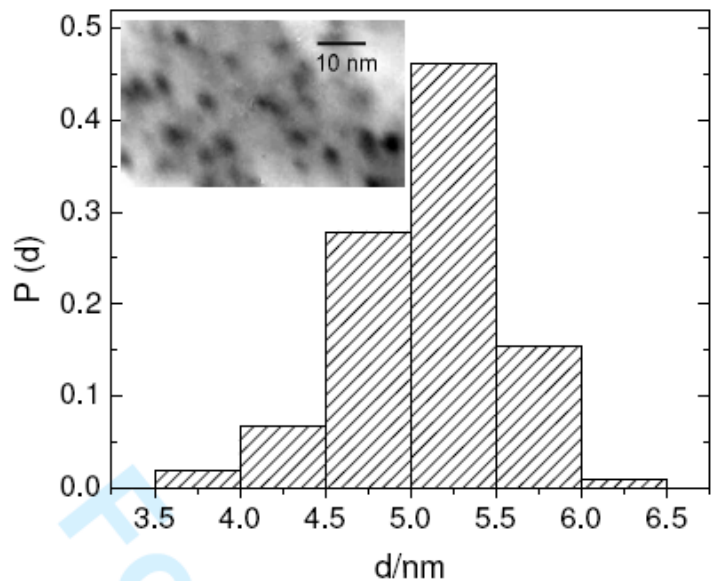

Figure 3. Histogram of the particle size distribution of $Cd_{0.5}Zn_{0.5}S$ nanocrystal. The inset shows the dark field TEM micrograph.

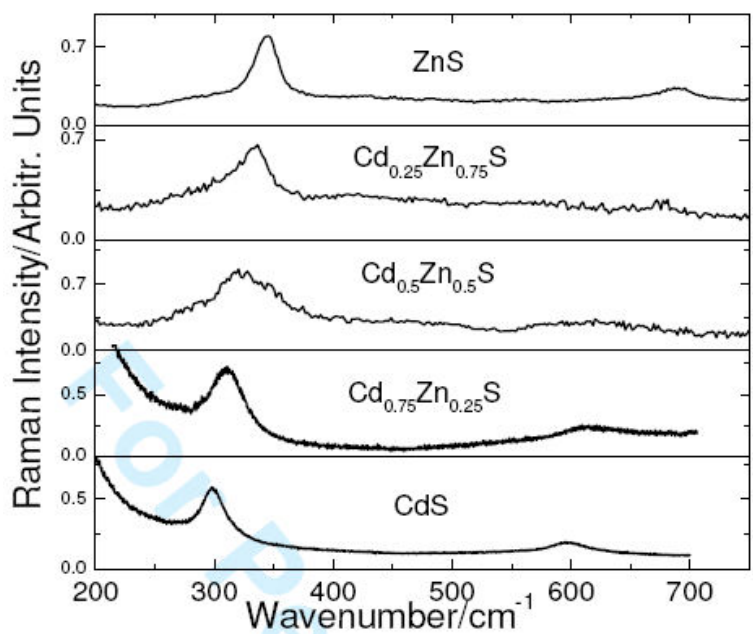

Figure 4. Optical phonon Raman spectra of $Cd_{1-x}Zn_xS$.

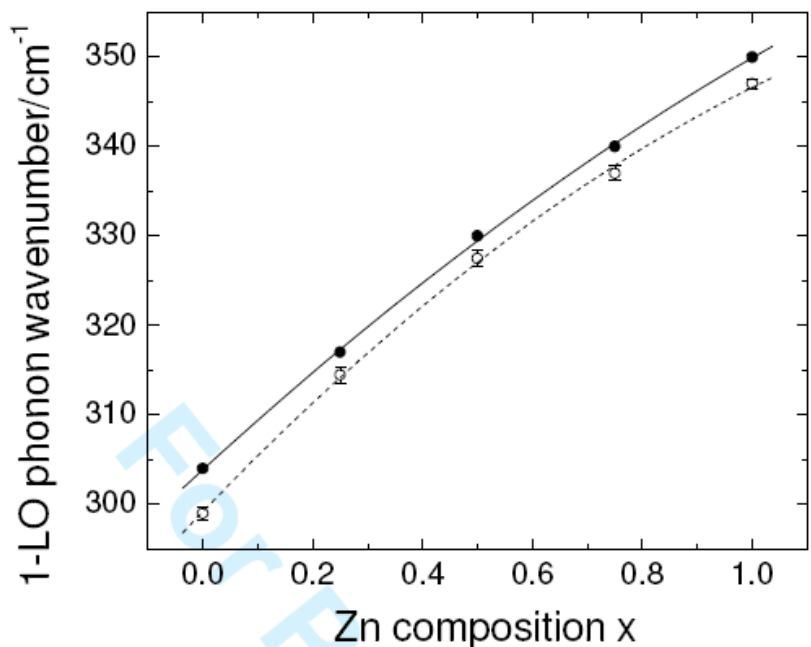

Figure 5. The variation of Raman peak positions with Zn compositions. Open symbols are the experimental data and filled symbols are the wavenumbers from MREI model for bulk $Cd_{1-x}Zn_xS$. The solid and dashed lines are guides to the eyes.

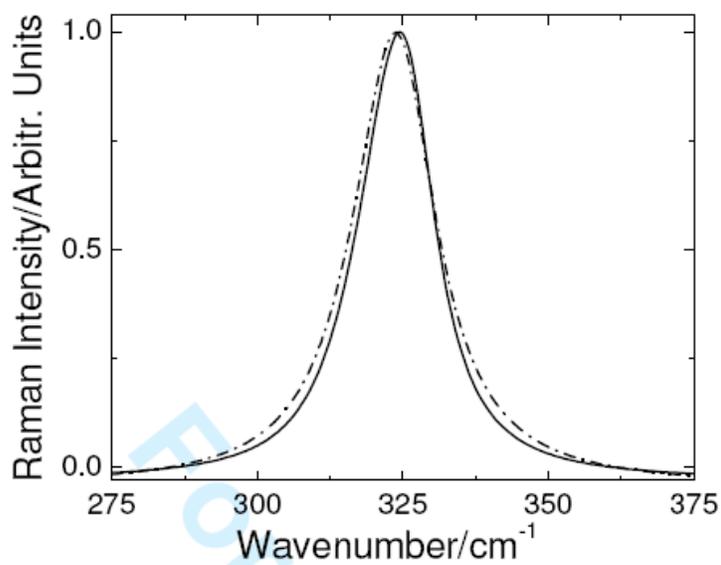

Figure 6. The comparison between the calculated Raman line-shapes obtained by taking the average particle size from XRD (dashed-dot curve) and the actual particle size distribution obtained from TEM (solid curve) for $Cd_{0.5}Zn_{0.5}S$ nanocrystal.

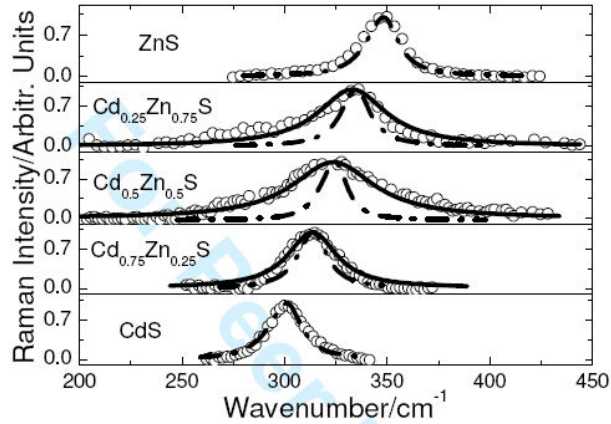

Figure 7. The comparison between calculated Raman line-shape and experimentally obtained spectra for all composition in $Cd_{1-x}Zn_xS$. Open symbols are the data. Dashed-dot curves are the calculated Raman line shape using phonon confinement model. Solid curves are the calculated line shape using $\Gamma'_0$ as the intrinsic linewidth in bulk mixed crystals.

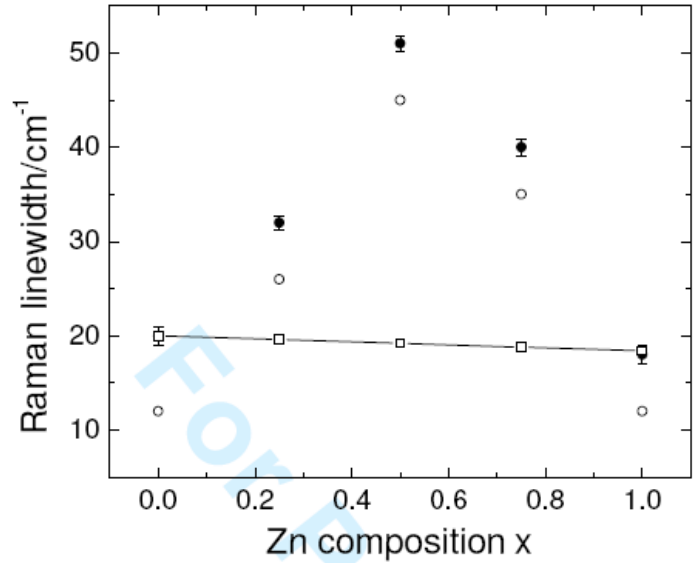

Figure 8. Variation of Raman line width with Zn composition. Filled symbols are experimentally obtained line widths and open symbols are the values of $\Gamma'_0$ that yield good agreement with measured spectra. Open squares represent the calculated Raman linewidth arising from phonon confinement for 3.2 nm nanocrystals.

A quantitative Raman line-shape analysis of confined optical phonons in $Cd_{1-x}Zn_xS$ mixed nanocrystals reveals that the substitutional disorder contributes much more to the line broadening than the phonon confinement effect. The contributions of the two mechanisms are also compared.

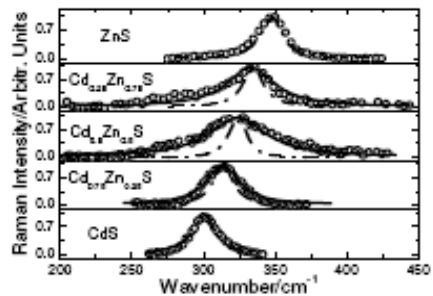

**Phonon confinement and substitutional disorder in $Cd_{1-x}Zn_xS$ Nanocrystals**

S. Sahoo*, S. Dhara, V. Sivasubramanian, S. Kalavathi and A. K. Arora